\documentstyle[12pt]{article}
\def\be {\begin{equation}}
\def\ee {\end{equation}}
\def\ba {\begin{eqnarray}}
\def\ea {\end{eqnarray}}
\newcommand{\bq}{\begin{eqnarray}}
\newcommand{\eq}{\end{eqnarray}}

\def\L  {\Lambda}

\def\bi {\begin{itemize}}
\def\ei {\end{itemize}}
\begin{document}
\def\bea{\begin{eqnarray}}
\def\eea{\end{eqnarray}}
\begin{center}
{\Large{\bf {Thermodynamical picture of the interacting holographic
dark energy model\\}}}
 \vskip 1cm
{\it Dedicated to the 50 year Jubilee of Professor Sergei D.
Odintsov}
\end{center}
\vspace{5mm}
\begin{center}
 {\large \bf M.R. Setare  \footnote{Some parts of work done in collaboration with E. C. Vagenas}
  \\{\it Department of Science,  Payame Noor University. Bijar. Iran}}
\date{\small{}}
\end{center}
\vspace{5mm} {\it Professor Sergei D. Odintsov  has been active in
various fields of theoretical physics, for this volume honouring his
contribution to theoretical physics, I have chosen a work
representative of our common interests.}

\begin{abstract}
In the present paper, we provide a thermodynamical interpretation
for the holographic dark energy model in a non-flat universe. For
this case, the characteristic length is no more the radius of the
event horizon ($R_E$) but the event horizon radius as measured from
the sphere of the horizon ($L$). Furthermore, when interaction
between the dark components of the holographic dark energy model in
the non-flat universe is present its thermodynamical interpretation
changes by a stable thermal fluctuation. A relation between the
interaction term of the dark components and this thermal fluctuation
is obtained.
 \end{abstract}

\newpage

\section{Introduction}
Nowadays it is strongly believed that the universe is experiencing
an accelerated expansion, and this is supported by many cosmological
observations, such as SNe Ia \cite{111}, WMAP \cite{2}, SDSS
\cite{3} and X-ray \cite{4}. These observations suggest hat the
universe is dominated by dark energy with negative pressure, which
provides the dynamical mechanism of the accelerating expansion of
the universe. Although the nature and origin of dark energy could
perhaps understood by a fundamental underlying theory unknown up to
now, physicists can still propose some paradigms to describe it. In
this direction we can consider theories of modified gravity
\cite{ordishov}, or field models of dark energy. The field models
that have been discussed widely in the literature consider a
cosmological constant \cite{cosmo}, a canonical scalar field
(quintessence) \cite{quint}, a phantom field, that is a scalar field
with a negative sign of the kinetic term \cite{phant,phantBigRip},
or the combination of quintessence and phantom in a unified model
named quintom \cite{quintom}. The quintom paradigm intends to
describe the crossing of the dark-energy equation-of-state parameter
$w_\L$ through the phantom divide $-1$ \cite{c14}, since in
quintessence and phantom models the perturbations could be unstable
as $w_\L$ approaches it \cite{c9}.

In addition, many theoretical studies are devoted to understand and
shed light on  dark energy, within the string theory framework. The
Kachru-Kallosh-Linde-Trivedi model \cite{kklt} is a typical example,
which tries to construct metastable de Sitter vacua in the light of
type IIB string theory. Despite the lack of a quantum theory of
gravity, we can still make some attempts to probe the nature of dark
energy according to some principles of quantum gravity. An
interesting attempt in this direction is the so-called ``holographic
dark energy'' proposal
\cite{Cohen:1998zx,Hsu:2004ri,Li:2004rb,holoext}. Such a paradigm
has been constructed in the light of holographic principle of
quantum gravity  \cite{holoprin}, and thus it presents some
interesting features of an underlying theory of dark energy.
Furthermore, it may simultaneously provide a solution to the
coincidence problem, i.e why matter and dark energy densities are
comparable today although they obey completely different equations
of motion \cite{Li:2004rb}. The holographic dark energy model has
been extended to include the spatial curvature contribution
\cite{nonflat} and it has been generalized in the braneworld
framework \cite{bulkhol}. Lastly, it has been tested and constrained
by various astronomical observations \cite{obs3}.\\
In the present paper we study the thermodynamical interpretation of
the interacting holographic dark energy model for a universe
enveloped by the event horizon measured from the sphere of the
horizon named $L$. We extend the thermodynamical picture in the case
where there is an interaction term between the dark components of
the HDE model. An expression for the interaction term in terms of a
thermal fluctuation is given. In the limiting case of flat
universe,we obtain the results derived in \cite{1}.

\section{ Intracting holographic dark energy density }
In this section we obtain the equation of state for the holographic
energy density when there is an interaction between holographic
energy density $\rho_{\Lambda}$ and a Cold Dark Matter(CDM) with
$w_{m}=0$. The continuity equations for dark energy and CDM are
\begin{eqnarray}
\label{2eq1}&& \dot{\rho}_{\rm \Lambda}+3H(1+w_{\rm \Lambda})\rho_{\rm \Lambda} =-Q, \\
\label{2eq2}&& \dot{\rho}_{\rm m}+3H\rho_{\rm m}=Q.
\end{eqnarray}
The interaction is given by the quantity $Q=\Gamma \rho_{\Lambda}$.
This is a decaying of the holographic energy component into CDM with
the decay rate $\Gamma$. the quantity $Q$ expresses the interaction
between the dark components. The interaction term $Q$ should be
positive, i.e. $Q>0$, which means that there is an energy transfer
from the dark energy to dark matter. The positivity of the
interaction term ensures that the second law of thermodynamics is
fulfilled \cite{Pavon:2007gt}.
At this point, it should be stressed that the continuity equations
imply that the interaction term should be a function of a quantity
with units of inverse of time (a first and natural choice can be the
Hubble factor $H$) multiplied  with the energy density. Therefore,
the interaction term could be in any of the following forms: (i)
$Q\propto H\rho_{X}$ \cite{Pavon:2005yx,Pavon:2007gt}, (ii)
$Q\propto H\rho_{m}$ \cite{Amendola:2006dg}, or (iii) $Q\propto
H(\rho_{X}+\rho_{m})$ \cite{Wang:2005ph}. The freedom of choosing
the specific form of the interaction term $Q$  stems from our
incognizance of the origin and nature of dark energy as well as dark
matter. Moreover, a microphysical model describing the interaction
between the dark components of the universe is not available
nowadays. Taking a ratio of two energy densities as $r=\rho_{\rm
m}/\rho_{\rm \Lambda}$, the above equations lead to
\begin{equation}
\label{2eq3} \dot{r}=3Hr\Big[w_{\rm \Lambda}+
\frac{1+r}{r}\frac{\Gamma}{3H}\Big]
\end{equation}
 Following Ref.\cite{Kim:2005at},
if we define
\begin{eqnarray}\label{eff}
w_\Lambda ^{\rm eff}=w_\Lambda+{{\Gamma}\over {3H}}\;, \qquad w_m
^{\rm eff}=-{1\over r}{{\Gamma}\over {3H}}\;.
\end{eqnarray}
Then, the continuity equations can be written in their standard
form
\begin{equation}
\dot{\rho}_\Lambda + 3H(1+w_\Lambda^{\rm eff})\rho_\Lambda =
0\;,\label{definew1}
\end{equation}
\begin{equation}
\dot{\rho}_m + 3H(1+w_m^{\rm eff})\rho_m = 0\; \label{definew2}
\end{equation}
We consider the non-flat Friedmann-Robertson-Walker universe with
line element
 \be\label{metr}
ds^{2}=-dt^{2}+a^{2}(t)(\frac{dr^2}{1-kr^2}+r^2d\Omega^{2}).
 \ee
where $k$ denotes the curvature of space k=0,1,-1 for flat, closed
and open universe respectively. A closed universe with a small
positive curvature ($\Omega_k\sim 0.01$) is compatible with
observations \cite{ {wmap}, {ws}}. We use the Friedmann equation to
relate the curvature of the universe to the energy density. The
first Friedmann equation is given by
\begin{equation}
\label{2eq7} H^2+\frac{kc^2}{a^2}=\frac{1}{3M^2_p}\Big[
 \rho_{\rm \Lambda}+\rho_{\rm m}\Big].
\end{equation}
Define as usual
\begin{equation} \label{2eq9} \Omega_{\rm
m}=\frac{\rho_{m}}{\rho_{cr}}=\frac{ \rho_{\rm
m}}{3M_p^2H^2},\hspace{1cm}\Omega_{\rm
\Lambda}=\frac{\rho_{\Lambda}}{\rho_{cr}}=\frac{ \rho_{\rm
\Lambda}}{3M^2_pH^2},\hspace{1cm}\Omega_{k}=\frac{kc^2}{a^2H^2}
\end{equation}
Now we can rewrite the first Friedmann equation as
\begin{equation} \label{2eq10} \Omega_{\rm m}+\Omega_{\rm
\Lambda}=1+\Omega_{k}.
\end{equation}
Using Eqs.(\ref{2eq9},\ref{2eq10}) we obtain following relation
for ratio of energy densities $r$ as
\begin{equation}\label{ratio}
r=\frac{1+\Omega_{k}-\Omega_{\Lambda}}{\Omega_{\Lambda}}
\end{equation}
In non-flat universe, our choice for holographic dark energy
density is
 \be \label{holoda}
  \rho_\Lambda=3c^2M_{p}^{2}L^{-2}.
 \ee
As it was mentioned, $c$ is a positive constant in holographic model
of dark energy($c\geq1$) and the coefficient 3 is for convenient.
$L$ is defined as the following form:
\begin{equation}\label{leq}
 L=ar(t),
\end{equation}
here, $a$, is scale factor and $r(t)$ can be obtained from the
following equation
\begin{equation}\label{rdef}
\int_0^{r(t)}\frac{dr}{\sqrt{1-kr^2}}=\int_t^\infty
\frac{dt}{a}=\frac{R_h}{a},
\end{equation}
where $R_h$ is event horizon. Therefore while $R_h$ is the radial
size of the event horizon measured in the $r$ direction, $L$ is the
radius of the event horizon measured on the sphere of the horizon.
 For closed universe we have (same calculation is valid for
open universe by transformation)
 \be \label{req}
 r(t)=\frac{1}{\sqrt{k}} sin y.
 \ee
where $y\equiv \sqrt{k}R_h/a$. Using definitions
$\Omega_{\Lambda}=\frac{\rho_{\Lambda}}{\rho_{cr}}$ and
$\rho_{cr}=3M_{p}^{2}H^2$, we get

\begin{equation}\label{hl}
HL=\frac{c}{\sqrt{\Omega_{\Lambda}}}
\end{equation}
Now using Eqs.(\ref{leq}, \ref{rdef}, \ref{req}, \ref{hl}), we
obtain
 \be \label{ldot}
 \dot L= HL+ a \dot{r}(t)=\frac{c}{\sqrt{\Omega_\Lambda}}-cos y,
\end{equation}
By considering  the definition of holographic energy density
$\rho_{\rm \Lambda}$, and using Eqs.( \ref{hl}, \ref{ldot}) one
can find:
\begin{equation}\label{roeq}
\dot{\rho_{\Lambda}}=-2H(1-\frac{\sqrt{\Omega_\Lambda}}{c}\cos
y)\rho_{\Lambda}
\end{equation}
Substitute this relation into Eq.(\ref{2eq1}) and using
definition $Q=\Gamma \rho_{\Lambda}$, we obtain
\begin{equation}\label{stateq}
w_{\rm \Lambda}=-(\frac{1}{3}+\frac{2\sqrt{\Omega_{\rm
\Lambda}}}{3c}\cos y+\frac{\Gamma}{3H}).
\end{equation}
Here as in Ref.\cite{WGA}, we choose the following relation for
decay rate
\begin{equation}\label{decayeq}
\Gamma=3b^2(1+r)H
\end{equation}
with  the coupling constant $b^2$. Using Eq.(\ref{ratio}), the
above decay rate take following form
\begin{equation}\label{decayeq2}
\Gamma=3b^2H\frac{(1+\Omega_{k})}{\Omega_{\Lambda}}
\end{equation}
Substitute this relation into Eq.(\ref{stateq}), one finds the
holographic energy equation of state
\begin{equation} \label{3eq4}
w_{\rm \Lambda}=-\frac{1}{3}-\frac{2\sqrt{\Omega_{\rm
\Lambda}}}{3c}\cos y-\frac{b^2(1+\Omega_{k})}{\Omega_{\rm \Lambda}}.
\end{equation}
\section{Thermodynamical interpretation of the interacting HDE model}
Following \cite{1} (see also \cite{Pavon:2007gt}), the
non-interacting HDE model in the non-flat universe as described
above is thermodynamically interpreted as a state in thermodynamical
equilibrium. According to the generalization of the black hole
thermodynamics to the thermodynamics of cosmological models, we have
taken the temperature of the event horizon to be $T_L=(1/2\pi L)$
which is actually the only temperature to handle in the system. If
the fluid temperature of the cosmological model is set equal to the
horizon temperature ($T_L$), then the system will be in equilibrium.
Another possibility \cite{davies2} is that the fluid temperature is
proportional to the horizon temperature, i.e. for the fluid
enveloped by the apparent horizon $T=eH/2\pi$ \cite{pavon}. In
general, the systems must interact for some length of time before
they can attain thermal equilibrium. In the case at hand, the
interaction certainly exists as any variation in the energy density
and/or pressure of the fluid will automatically induce a
modification of the horizon radius via Einstein's equations.
Moreover, if $T \neq T_{L}$, then energy would spontaneously flow
between the horizon and the fluid (or viceversa), something at
variance with the FRW geometry \cite{pa}. Thus, when we consider the
thermal equilibrium state of the universe, the temperature of the
universe is associated with the horizon temperature. In this picture
the equilibrium entropy of the holographic dark energy is connected
with its energy and pressure through the first thermodynamical law
\be \label{law1} TdS_{\Lambda}=dE_{\Lambda}+p_{\Lambda}dV \ee where
the volume is given as \be V=\frac{4\pi}{3}L^{3} \hspace{1ex}, \ee
the energy of the holographic dark energy is defined as \be
\label{energy1} E_{\Lambda}=\rho_{X} V=4\pi c^{2} M^{2}_{p}L \ee and
the temperature of the event horizon is given as \be \label{temp1}
T=\frac{1}{2\pi L^{0}} \hspace{1ex}. \ee Substituting the aforesaid
expressions for the volume, energy, and temperature in equation
(\ref{law1}) for the case of the non-interacting HDE model, one
obtains \be \label{entropy1}
dS_{\Lambda}^{(0)}=8\pi^{2}c^{2}M^{2}_{p}\left(1+3\omega^{0}_{\Lambda}\right)L^{0}dL^{0}
\ee where the superscript $(0)$ denotes that in this thermodynamical
picture our universe is in a thermodynamical stable equilibrium.\\
In the interacting case, by substituting equation (\ref{roeq}) in
the conservation equation (\ref{2eq1}) for the dark energy component
one obtains \be \label{eosp2}
1+3\omega_{\Lambda}=-2\frac{\sqrt{\Omega_{\Lambda}}}{c}\cos y -
\frac{Q}{3H^{3}M^{2}_{p}\Omega_{\Lambda}} \hspace{1ex}. \ee
According to \cite{1}, the interacting HDE model in the non-flat
universe as described above is not anymore thermodynamically
interpreted as a state in thermodynamical equilibrium. In this
picture the effect of interaction between the dark components of the
HDE model is thermodynamically interpreted as a small fluctuation
around the thermal equilibrium. Therefore, the entropy of the
interacting holographic dark energy is connected with its energy and
pressure through the first thermodynamical law \be \label{law2}
TdS_{\Lambda}=dE_{\Lambda}+p_{\Lambda}dV \ee where now the entropy
has been assigned an extra logarithmic correction \cite{das} \be
S_{\Lambda}=S_{\Lambda}^{(0)}+S_{\Lambda}^{(1)} \ee where \be
\label{correction1} S_{\Lambda}^{(1)}=-\frac{1}{2}\ln \left(C
T^{2}\right) \ee and $C$ is the heat capacity defined by \be
C=T\frac{\partial S_{\Lambda}^{(0)}}{\partial T} \ee and using
equations (\ref{entropy1}), (\ref{temp1}), is given as \bea
\label{capacity1}
C&\hspace{-1ex}=\hspace{-1ex}&-8\pi^{2}c^{2}M^{2}_{p}(L^{0})^{2}(1+3\omega_{\Lambda}^{0})\\
\label{capacity2} &\hspace{-1ex}=\hspace{-1ex}&16\pi^{2}c
M^{2}_{p}(L^{0})^{2} \sqrt{\Omega_{\Lambda}^{0}} \cos y
\hspace{1ex}. \eea Substituting the expressions for the volume,
energy, and temperature (it is noteworthy that these quantities
depend now on $L$ and not on $L^{0}$ since there is interaction
among the dark components) in equation (\ref{law2}) for the case of
the interacting HDE model, one obtains \be \label{entropy3}
dS_{\Lambda}=8\pi^{2}c^{2}M^{2}_{p}\left(1+3\omega_{\Lambda}\right)L
dL \ee and thus one gets \bea \label{eosp3}
1+3\omega_{\Lambda}&\hspace{-1ex}=\hspace{-1ex}&\frac{1}{8\pi^{2}c^{2}M^{2}_{p}L}\frac{dS_{\Lambda}}{dL}\\
&\hspace{-1ex}=\hspace{-1ex}&\frac{1}{8\pi^{2}c^{2}M^{2}_{p}L}\left[\frac{dS_{\Lambda}^{(0)}}{dL}+\frac{dS_{\Lambda}^{(1)}}{dL}\right]\\
\label{eosp4}
&\hspace{-1ex}=\hspace{-1ex}&-2\left(\frac{\sqrt{\Omega_{\Lambda}^{0}}}{c}\cos
y \right)\frac{L^{0}}{L}
\frac{dL^{0}}{dL}+\frac{1}{8\pi^{2}c^{2}M^{2}_{p}L}\frac{dS_{\Lambda}^{(1)}}{dL}
\eea where the last term concerning the logarithmic correction can
be computed using expressions (\ref{correction1}) and
(\ref{capacity2}) \be
\frac{dS_{\Lambda}^{(1)}}{dL}=-\frac{H}{\left(\frac{c}{\sqrt{\Omega_{\Lambda}^{0}}}-\cos
y\right)}
\left[\frac{\left(\Omega_{\Lambda}^{0}\right)'}{4\Omega_{\Lambda}^{0}}+
y \tan y\right] \ee with the prime $(\hspace{1ex}')$ to denote
differentiation with respect to $\ln a$.
\par\noindent
Therefore, by equating the expressions (\ref{eosp2}) and
(\ref{eosp4}) for the equation of state parameter of the holographic
dark energy evaluated on cosmological and thermodynamical grounds
respectively, one gets an expression for the interaction term \be
\label{interaction} \frac{Q}{9H^{3}M^{2}_{p}}=
\frac{\Omega_{\Lambda}}{3}\left[-\frac{2\sqrt{\Omega_{\Lambda}}}{c}\cos
y +\left(\frac{2\sqrt{\Omega_{\Lambda}}}{c}\cos
y\right)\frac{L^{0}}{L}\frac{dL^{0}}{dL}\right]-
\frac{1}{8\pi^{2}c^{2}M^{2}_{p}L}\frac{\Omega_{X}}{3}\frac{dS_{\Lambda}^{(1)}}{dL}
\hspace{1ex}. \ee It is noteworthy that in the limiting case of flat
universe, i.e. $k=0$, we obtain exactly the result derived in
\cite{1} when one replaces $L^0$ and $L$ with $R^{0}_E$ and $R_E$,
respectively.
 \section{Conclusions}
It is of interest to remark that in the literature, the different
scenarios of DE has never been studied via considering special
similar horizon, as in \cite{davies2} the apparent horizon, $1/H$,
determines our universe. For flat universe the convenient horizon
looks to be event horizon, while in non flat universe we define $L$
because of the problems that arise if we consider event horizon or
particle horizon (these problems arise if we consider them as the
system's IR cut-off). Thus it looks that we need to define a horizon
that satisfies all of our accepted principles; in \cite{odintsov} a
linear combination of event and apparent horizon, as IR cut-off has
been considered. In present paper, we studied $L$, as the horizon
measured from the sphere of the horizon as system's IR cut-off. In
the present paper, we have provided a thermodynamical interpretation
for the HDE model in a non-flat universe. We utilized the horizon's
radius $L$ measured from the sphere of the horizon as the system's
IR cut-off. We investigated the thermodynamical picture of the
interacting HDE model for a non-flat universe enveloped by this
horizon. The non-interacting HDE model in a non-flat universe was
thermodynamically interpreted as a thermal equilibrium state. When
an interaction between the dark components of the HDE model in the
non-flat universe was introduced the thermodynamical interpretation
of the HDE model changed. The thermal equilibrium state was
perturbed by a stable thermal fluctuation which was now the
 thermodynamical interpretation
of the interaction. Finally, we have derived an expression that
connects this interaction term of the dark components of the
interacting HDE model in a non-flat universe with the aforesaid
thermal fluctuation.
\section{Acknowledgment}
I deeply appreciate the invitation of Professor P. M. Lavrov and
Professor V. Ya. Epp to submit this article to the anniversary
volume "The Problems of Modern Cosmology" on the occasion of the
$50$th birthday of Professor Sergei D. Odintsov.

\end{document}